\documentclass[pdflatex,sn-mathphys]{sn-jnl}
 
\jyear{2022}%

\theoremstyle{thmstyleone}%
\theoremstyle{thmstyletwo}%

\theoremstyle{thmstylethree}%

\usepackage{listings}
\usepackage{algpseudocode}
\algnewcommand\algorithmicnot{\textbf{not}}
\algdef{SE}[IF]{IfNot}{EndIf}[1]{\algorithmicif\ \algorithmicnot\ #1\ \algorithmicthen}{\algorithmicend\ \algorithmicif}%
\usepackage[section]{placeins}
\usepackage{graphicx}
\usepackage[justification=centering]{caption}

\begin{document}

\title[BigBird: Big Data Storage and Analytics at Scale in Hybrid Cloud]{BigBird: Big Data Storage and Analytics at Scale in Hybrid Cloud}


\author*[1]{\fnm{Saurabh} \sur{Deochake}}\email{sdeochake@twitter.com}

\author[2]{\fnm{Vrushali} \sur{Channapattan}}\email{vchannapattan@twitter.com}
\equalcont{These authors contributed equally to this work.}

\author[3]{\fnm{Gary} \sur{Steelman}}\email{gsteelman@twitter.com}
\equalcont{These authors contributed equally to this work.}

\affil*[1]{\orgname{Twitter Inc.}, \orgaddress{\street{1355 Market St \#900}, \city{San Francisco}, \postcode{94103}, \state{California}, \country{United States of America}}}


\abstract{Implementing big data storage at scale is a complex and arduous task that requires an advanced infrastructure. With the rise of public cloud computing, various big data management services can be readily leveraged. As a critical part of Twitter's ``Project Partly Cloudy", the cold storage data and analytics systems are being moved to the public cloud. This paper showcases our approach in designing a scalable big data storage and analytics management framework using BigQuery in Google Cloud Platform while ensuring the security, privacy and data protection. The paper also discusses the limitations on the public cloud resources and how they can be effectively overcome when designing a big data storage and analytics solution at scale. Although the paper discusses the framework implementation in Google Cloud Platform, it can easily be applied to all major cloud providers.}




\maketitle

\section{Introduction}\label{sec1}
Every day, over a hundred million people come to Twitter to find out what’s happening in the world and have conversations about it. Every Tweet, Retweet, reply and any other user action generates an event that we store and make available for analytics at Twitter. This user-generated data is stored in Twitter’s real-time and ad-hoc processing Hadoop clusters. Our Hadoop infrastructure consists of tens of thousands of servers comprising dozens of clusters \cite{bib1}. As Twitter users interact with our services, they generate events that are collected through our log pipeline infrastructure and distributed to downstream services.

While Twitter had cost-effective on-premise data management infrastructure, there was a desire and motivation to extend these capabilities to the public cloud. There is a multitude of advantages of moving to cloud viz. elasticity, scalability, ability to scale over a broader geographical footprint for the business continuity, and leveraging new cloud service offerings and capabilities as they become available \cite{bib2}. Twitter’s Partly Cloudy is a project to extend data processing at Twitter from an on-premise only model to a hybrid cloud model on Google Cloud Platform (GCP). As a part of this project, Twitter migrated its ad-hoc and cold storage Hadoop data processing clusters to GCP. To continue the data processing in the cloud, Twitter also had to replicate over 300 PB of data from on-premise HDFS storage systems to Google Cloud Storage (GCS) \cite{bib3}.

As we migrated our data to GCS, we quickly realized a need for the scalable big data storage and analytics solution for users and internal teams at Twitter. Having a data warehouse solution in the cloud will empower these users and teams to analyze and visualize the data to improve how new features are built on Twitter. We identified GCP Big Data warehouse and analytics services as having the potential to help our big data management, analytics, and machine learning initiatives at Twitter. Project BigBird uses GCP BigQuery \cite{bib4}, an enterprise data warehouse with a fast SQL engine based on Dremel \cite{bib5}, capable of providing machine learning solutions, to store over hundreds of petabytes of data. 

This paper showcases the big data storage management framework that Twitter designed along with the solutions for data security, access control, and big data operations. We hope that Project BigBird can act as a case study of storing a massive amount of data in the public cloud for our industry peers as well as the big data research community. The paper is structured as follows: Section \ref{sec2} discusses the previous work of the Partly Cloudy project in detail, section \ref{sec3} showcases our big data storage hierarchy framework using BigQuery, section \ref{sec4} presents the operations (ops) side of our approach, and finally, we conclude this paper in section \ref{sec5}.

\section{Related Work}\label{sec2}
\begin{figure*}
  \centering\includegraphics[width=0.4\linewidth]{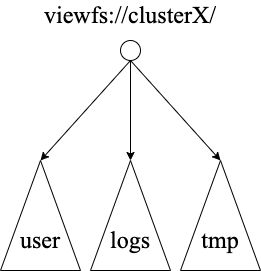}
  \caption{Twitter's HDFS ViewFS Namespaces}
  \label{fig:viewfs}
\end{figure*}

Twitter's Hadoop storage systems store more than 300 PB of data across tens of thousands of servers. HDFS Federation aids us in scaling our HDFS file system and sustaining high HDFS object counts. While this approach is great for scaling, it is not easy to use since each member namespace in the federation has its own URI. Therefore, as shown in Fig. \ref{fig:viewfs}, we use TwitterViewFS \cite{bib1}, an extension to ViewFS, to provide an illusion of a single namespace in a cluster by creating a mount table with links to namespaces like user and logs. Here, the user is either a Twitter employee user or a service account that owns a specific Twitter service that generates the HDFS storage data. On the other hand, logs are a type of data that are generated by Twitter users when they interact with each other and Twitter, the platform itself. Now, with the help of TwitterViewFS, we can easily map our Hadoop clusters like \textit{/dc1/cluster1/user/helen}, where dc1 and cluster1 represent a particular data center and a Hadoop cluster, respectively.

\subsection{Project Partly Cloudy}{\label{section:partly-cloudy}}
Partly Cloudy project is a pivotal step in replicating the on-premise HDFS data to Google Cloud Storage (GCS). While we had the HDFS file system path for the data, we did not have the corresponding GCS buckets. To achieve that resource parity in the cloud, we developed an in-house suite of services called Demigod. The Demigod suite of services are responsible for a) managing the life cycle of the GCS buckets corresponding to the on-premise HDFS file path b) managing the life cycle of the ``shadow" service accounts in GCP that correspond to the on-premise UNIX users and c) assigning the appropriate permissions for ``shadow" accounts and users to these buckets so that the services can read and write data to the buckets. Section \ref{sec:user_auth} discusses the concept of ``shadow" accounts. The working of Demigod services can be a different topic to discuss and therefore, not in scope for this paper. Since all buckets in GCS live in a global namespace, to avoid bucket names clash with other GCP users, twitter.domain is a verified bucket domain name used for Twitter's big data storage in GCS. Therefore, for uniqueness, GCS buckets in Twitter's GCP would end with a ``twitter.domain" string. Now that we had GCS resources in GCP, our ``replicator" services needed to be aware of the path conversion from HDFS to GCS. Therefore, we implemented a path conversion mechanism to logically map an on-premise HDFS path to a globally unique GCS bucket path. This conversion happens in two stages. In the first stage, the on-premise HDFS path is converted into a logical GCS path, for example, \textit{/dc1/cluster1/user/helen/some/path/part-001.lzo} is converted into logical GCS path \textit{gcs/user/helen/some/path/part-001.lzo}. Then in the second stage, the logical path is converted into an actual GCS bucket path viz. \textit{gs://user.helen.dp.twitter.domain/some/path/part-001.lzo}. Similarly, for logs, the GCS path may be \textit{gs://log.activity-logs.dp.twitter.domain/some/path/part-001.lzo}. We provision these GCS buckets under a dedicated GCP project named like ``twitter-gcs-project". This path conversion methodology helps the replication services seamlessly find a GCS bucket path for a corresponding HDFS file system path. These paths form the basis of data replication and ingestion from on-premise HDFS file system into GCS, and eventually from GCS into BigQuery.

\subsubsection{Data Replication}
To make data available in GCS, we built an in-house service called the Data Replicator Service that aids in copying the on-premise HDFS data to GCS \cite{bib6}. To achieve the data replication with minimal impact on the production Hadoop clusters, we employ a dedicated Copy Cluster per data center. The intended purpose of the Copy Cluster is to get the data for the input path from the production cluster and copy it to GCS over a private link to GCP. The mapping between the HDFS path to the logical GCS path, as explained above, makes this replication transparent to the jobs running on these Copy Clusters. The biggest benefit of using a Copy Cluster is that the jobs running on these clusters do not consume precious CPU time on the ad-hoc clusters. Once the data is copied over to GCS, multiple services, including ad-hoc Hadoop clusters running in GCP, can make use of this data for various purposes like data analytics and machine learning.

\begin{figure}[!htbp]
  \centering\includegraphics[width=\linewidth]{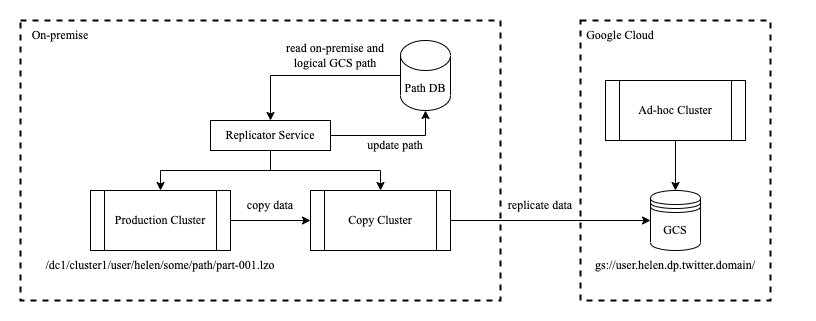}
  \caption{Data Replication from HDFS to Google Cloud Storage (GCS)}
  \label{fig:auth}
\end{figure}

\subsubsection{User Authentication}\label{sec:user_auth}
A crucial part of enabling big data storage and management on GCP is how users authenticate to use the resources. Like the rest of the industry, Twitter uses LDAP \cite{bib7} and UNIX-based user identities on-premise. Additionally, the users also get GSuite, also known as Google Workspace, identity when they log into GCP. Unfortunately, GCP does not understand UNIX-based identity systems. Therefore, we needed a way to bridge the gap between the UNIX-based identities and GSuite-based identities that Google provisions to enable programmatic data access and operations. We implemented a ``shadow" identity in GCP that is tied to the on-premise identity of a user. The shadow identity is a GCP Service Account created for an on-premise user for programmatic access to the data stored in public cloud. Therefore, as showcased in Fig. \ref{fig:auth}, an on-premise user with UNIX identity as ``helen" will have GSuite identity as ``helen@gsuite.domain" for direct interaction with their GCS bucket and a shadow service account named ``helen@gserviceaccount.com`` for programmatic access to the GCS bucket. The user's UNIX identity will have access to the Hadoop clusters via authentication protocols like Kerberos \cite{bib8}. The Demigod suite of services, as discussed in the above sections, creates the shadow service account identity and applies appropriate access control for this identity on the GCS buckets. Our proposed BigQuery-based architecture also follows a similar model for access control as discussed in the latter sections of this paper.

\begin{figure}[!htbp]
  \centering\includegraphics[width=0.7\linewidth]{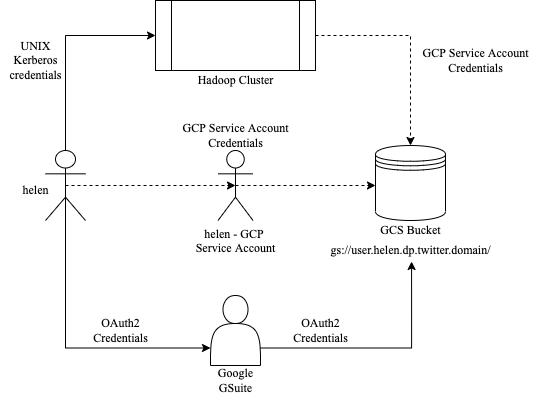}
  \caption{User Authentication in Google Cloud}
  \label{fig:auth}
\end{figure}

\FloatBarrier

\subsubsection{Overall Partly Cloudy GCS Workflow}
This section showcases the steps that are performed by Demigod services in creating and managing the resources in GCP. It is imperative to understand the overall workflow of GCS resource creation since our framework extends this workflow to BigQuery.

\paragraph{Creation of Shadow Service Account}
For the on-premise UNIX user, a shadow service account with the appropriate name is created by invoking GCP REST APIs. Additionally, the secret key for this service account is securely stored in a key storage vault. This secret key is rotated after every few days thereby reducing the risk of exposure to attacks.

\paragraph{Creation of GCS Bucket}
Once the shadow service account and its key are successfully created for a user, the Demigod services create bucket for the user with a globally unique name showcased in the section above. These buckets are created under a common GCP project named ``twitter-gcs-project", for example. A GCP project is an entity that organizes all the GCP resources along with their configurations inside a theoretical container.

\paragraph{Set up Access Control}
For the user to access their GCS data seamlessly, the Demigod services also applies a few permissions on the buckets, abiding by the  Authentication, Authorization, and Accounting (AAA) principle \cite{bib9} and the Principle of Least Privilege \cite{bib10}. The user's GSuite identity and their shadow service account identity get ``owner" permissions on the GCS bucket. Additionally, the service also creates a ``reader" group with read-only permissions where the user can share their data with other users by requesting them to join the ``reader" group. The group, mentioned here, is a Google group. 

\section{Our Framework}\label{sec3}
Before we discuss our big data management framework using BigQuery, we must visit a few concepts pertaining to GCP. We introduce them in brief as follows:

\begin{description}
  \item[$\bullet$ \textbf{Organization}]: An Organization is a root node of the GCP resource hierarchy. 
  \item[$\bullet$ \textbf{Folder}]: Folders are nodes that group similar resources like projects, folders, or a combination of both.
  \item[$\bullet$ \textbf{Project}]: Projects form the basis of creating, managing and enabling all GCP services and APIs.
  \item[$\bullet$ \textbf{GCS Bucket}]: Buckets are basic storage containers that hold your data.
  \item[$\bullet$ \textbf{BigQuery Dataset}]: Datasets are top-level containers that organize and control access to the data stored inside a table.
  \item[$\bullet$ \textbf{BigQuery Table}]: Tables are basic forms of data storage. A table contains individual records organized in rows.
  \item[$\bullet$ \textbf{BigQuery View}]: A View is a virtual table that acts as a result set of a stored query on the data
  \item[$\bullet$ \textbf{BigQuery Slot}]: A Slot is a unit of virtual CPU used by BigQuery to execute SQL queries.
\end{description}

\begin{figure*}
  \centering\includegraphics[width=\linewidth]{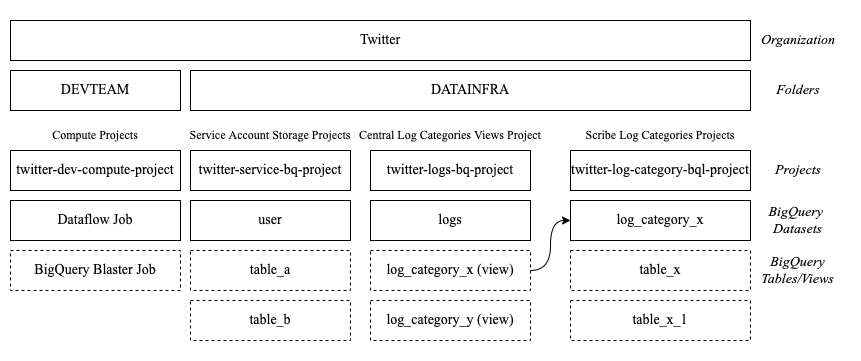}
  \caption{BigBird Resource Hierarchy}
  \label{fig:bq_org}
\end{figure*}

\noindent This section showcases how Twitter manages hundreds of petabytes of data in GCP BigQuery. The section is divided into five parts. Firstly, we discuss the limitations with the initial proposal based on a centralized data storage system based on the Partly Cloudy project's GCS-based implementation, Secondly, the section showcases how the data can be partitioned and stored in different GCP projects. Thirdly, we explain how data is replicated from non-BigQuery data sources like HDFS and GCS into BigQuery. Subsequently, we illustrate the computing needs for BigQuery Analytics. Finally, we demonstrate our logging and monitoring solutions that aid in scalable BigQuery operations.

\subsection{Limitations of Initial Proposal}
When we started working on enabling big data storage and analytics at Twitter, we wanted to keep the user experience similar to how they interact with data stored in GCS. This would mean that we would have a single GCP project that stores all BigQuery datasets for the users and services, an approach that was similar to the data storage framework using GCS as mentioned in the above sections. Although this would have made it easier for services and users to access BigQuery data, we quickly realized that the scale of our data and operations would not fit quotas and limitations GCP has put in place for BigQuery \cite{bib11}. Every service in GCP has associated quotas and limitations in place for scalability purposes. 

\subsubsection{Projects}
Our initial design proposed having a single project that would store all the datasets and tables for all service accounts and logs. However, there are a few stark limitations on BigQuery usage in a GCP project viz., a) there may be up to 1000 datasets after which the performance degradation via CLI or API is significant, b) BigQuery streaming API limits requests per second (RPS) to 10M/s per project. This means that this rate would be too small for the scale of our usage, c) Data Transfer Service executes data load jobs and counts against project level quotas for load jobs thereby making it unfit for our scale.

\subsubsection{Datasets}
Datasets are our fundamental storage solution to store our data in our framework. In the initial proposal, we planned to create all datasets for service accounts and logs inside a single project. However, as explained above, the performance degradation for CLI and APIs after 1000 datasets is immense. Additionally, after 50000 tables inside a dataset, we encounter similar performance degradation making this design unusable at scale. Finally, we could store only 2500 authorized views in a single dataset, a small number for monolithic design.

\subsubsection{Tables}
Similar to datasets, tables had a few important limitations for our initial design viz., a) performance degradation after 50000 tables, b) 1000 operations per table per day, including failures, c) only 5 metadata operations every 10 seconds per table, and d) a table may have only maximum of 10000 columns. These limitations clearly could not handle our scale if we had all services and users query data stored in the tables in the same project.

\subsubsection{Data Loading}
Data loading and unloading is an essential part of our framework since data is loaded from GCS to BigQuery for performing analytics and machine learning. The limitations for batch load jobs were 50 concurrent load jobs per 2000 slots before job queuing is engaged by Google and 100000 load jobs per project per day. The limits were too strict for our scale and use case.

\subsubsection{API Limits}
For programmatic access to BigQuery, a service can use BigQuery Storage APIs to read from tables or write data to a table. When a service or user reads data using the BigQuery Storage API via the \texttt{ReadRows} API call, the end-user is limited to only 5000 \texttt{ReadRows} calls per minute, per user, per project. Additionally, for all other API calls, the users are limited to 1000 BigQuery Storage API calls per minute, per user, per project. \\

\noindent The limitations showcased were too stringent for running a truly scalable big data storage and analytics platform, especially if all datasets were created under one parent project like the GCS framework. Therefore, we implemented a structured and scalable framework that would solve all the limitations with the initial proposal while ensuring that the user experience is as simple as possible.

\begin{algorithm}
\caption{Precondition BigBird Resource Creation}\label{alg:precondition}
\begin{algorithmic}
\Procedure{Precondition}{}
\State System Initialization
\State Read List of Users and Logs 
\For{\textit{each user}} \Comment{Repeat for Logs}
\IfNot{\textit{user.hasShadowAccount}}
\State Create Shadow Service Account
\EndIf
\IfNot{\textit{user.hasGCSBucket}}
\State Create GCS Bucket
\State Assign Ownership Permission on Bucket to Shadow Account
\If {\textit{read-only Google group does not exist}}
\State Create New Google group
\Else
\State Assign Read-only Permission on Bucket to Reader Group
\EndIf
\EndIf
\If {\textit{precondition successful}}
\State \Return True
\Else
\State \Return False
\EndIf
\EndFor
\EndProcedure
\end{algorithmic}
\end{algorithm}

\begin{algorithm}
\caption{Provision BigBird Big Data Resources}\label{alg:provision}
\begin{algorithmic}
\Procedure{ProvisionBigBird}{}
\State System Initialization
\State Read List of Users and Logs 
\For{\textit{each user}} 
\State $status\gets\textsc{Precondition}(user)$
\If{status = False}
\State \Return Failure
\Else
\State Create GCP Project - ``twitter-user-bq-project"
\State Apply Labels to Identify as Big Data Storage Project
\State Create BigQuery Dataset - ``user"
\State Assign BigQuery Ownership Permission to User Shadow Account
\State Assign BigQuery Read-only Permission to Reader Group
\EndIf
\EndFor
\For{\textit{each log category}}
\State $status\gets\textsc{Precondition}(user)$
\If{status = False}
\State \Return Failure
\Else
\State Create GCP Project - ``twitter-log-bql-project" \Comment{``bql" for logs}
\State Create View Under ``logs" Dataset in ``twitter-logs-bq-project"
\State Assign BigQuery Read-only Permission to Reader Group for View
\State Create BigQuery Dataset in ``twitter-log-bql-project"
\State Assign BigQuery Ownership Permission to Data Replication Shadow Account
\State Assign BigQuery Read-only Permission to Reader Group for Log Dataset
\EndIf
\EndFor
\EndProcedure
\end{algorithmic}
\end{algorithm}

\subsection{Data Storage Framework}
In our framework design, GCP Projects are a fundamental way to organize and access data. Our design recommends separating storage and compute for security adherence, resource cost chargeback, and manageability. The design principles in separating storage and compute along with human and machine accounts were largely influenced by Twitter’s compliance needs. Using this recommended project architecture for storing data provides an easy way for data registration, annotation, retention, and wipeout that would otherwise need to be solved separately by individual users and teams.

\subsubsection{Compute Projects}
Compute projects are specially designed to run an ad-hoc query or a compute job, such as GCP Dataflow, to read and write data to BigQuery Storage projects. Each compute project is owned by an individual team where compute resources like BigQuery slots and access to run queries are shared by an entire team. The cost incurred in performing the computing operations is automatically charged to the team via the Cloud Chargeback Service.

\subsubsection{Storage Projects}
Storage projects are a crucial part of our design framework. These projects securely store the data for a user and log category. Algorithm \ref{alg:provision} showcases the workflow that provisions the projects and datasets that store the data. The automation, as mentioned in algorithm \ref{alg:precondition} service reads the list of users (also referred to as Service Account users) and Hadoop Scribe log categories from on-premise and it performs preconditioning on each user and log category where it ensures that the identity and GCS bucket infrastructure is set up for them. If the preconditioning is successful then the automation service creates service account user and log categories projects and resources. Following sections deliberate on these projects in detail.

\paragraph{Service Account User Projects}
Service accounts are on-premise HDFS users that store the data for a particular Twitter service. In our design framework, we create a new GCP project for each service account user that fulfills the preconditioning requirements. When the preconditioning is successful, the automation service creates a new project named ``twttr-user-bq-project" with an empty BigQuery dataset named ``user" to signify that the project stores service account user data just like the GCS bucket naming as mentioned in section \ref{section:partly-cloudy}. Due to Twitter's data compliance requirements, it is imperative that the service account user projects follow a least privilege access principle. Therefore, we apply an ownership role for the service account user's shadow account in GCP at the ``user" dataset. This enables creating new tables, writing data, and editing the schema possible for the service account user. Furthermore, to facilitate the seamless reading of data stored inside this dataset, we create a new Google group called the reader group and assign it read-only permission. By designing a read-only group, we ensure that the data can be read by other authorized users but cannot be edited. 

\paragraph{Scribe Log Categories Projects}
Log categories at the aggregation of real-time logs that are generated when Twitter users interact with the services. These logs are then categorized into a multitude of categories, for example, a ``system-event" category or a ``user-event" category. Since the Scribe service is used to aggregate these logs, we will refer to log projects are Scribe Log Categories projects. Unlike service account user projects, our framework employs a different technique to store the logs data in GCP. As mentioned in detail in algorithm \ref{alg:provision}, when the preconditioning is successful for a log category, we create a new project named like ``twitter-log-bql-project". To differentiate between service account user and log category project, we add `bql' identifier in the project name. Subsequently, a new dataset named exactly like the log category name is created in the project. Since log categories generate aggregated data based on the interaction of multiple services, when replicating this data to GCP, we designate special shadow service accounts for the ``Replication Services". We will discuss in detail how replication works in our framework in the next subsection. Our automation service assigns an ownership role for the replication service's shadow account. We also create a reader group for this log category dataset and assign read-only permission to it. Furthermore, since log categories ownership attribution is complex due to the nature of aggregation of data owned by different services, to make the user experience easier, we create a centralized project named ``twitter-logs-bq-project" that stores the BigQuery Views to actual log categories data stored in individual log categories project. These BigQuery views to the actual datasets stored in a ``bql" project also get a reader group with read-only permissions. This ensures that views are designed to hide the backend implementation details for a Scribe log category.

\subsection{Data Ingestion}
There are a few ways the services and users can ingest their data from HDFS or GCS into BigQuery. Our framework provides flexibility in ingesting the data into BigQuery including the scale of ingestion and ease of operations. We explain a few of major ways of ingestion as follows and compare them in Table \ref{tab:compare-tools}. For the purpose of ingesting the data, our framework designates a dedicated ``data load" project named ``twitter-gcs-to-bq-project" and data is loaded from the GCS bucket into corresponding BigQuery dataset via this project.

\begin{figure*}
  \centering\includegraphics[width=\linewidth]{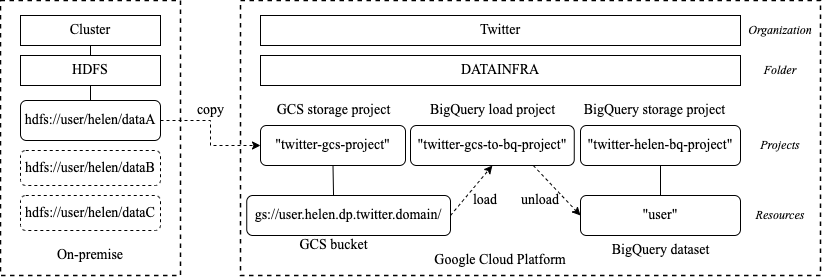}
  \caption{BigQuery Data Ingestion}
  \label{fig:ingestion}
\end{figure*}

\subsubsection{BigQuery Blaster}
BigQuery Blaster is a set of in-house libraries that enable services and users to load typed data sources into BigQuery via Beam or Hadoop. Additionally, using the Scalding library, users can create transfer jobs that read the data from the corresponding GCS bucket via a \texttt{TypedPipe} and write to BigQuery through the \texttt{BigQuerySink} data sink option. Furthermore, using the Apache Beam library, users can use the same BigQuery Blaster logic as stated above to write \texttt{thrift structs} to BigQuery via \texttt{BigQueryIO} option.

\subsubsection{BigQuery Ingestion with Data Transfer Service (BID)}
BigQuery Ingestion using Data Transfer Service (BID) is another in-house ingestion framework that runs on top of Google Kubernetes Engine (GKE) and uses BigQuery Data Transfer Service. BID consumes a user-generated input with the source path of GCS bucket and destination path for their corresponding BigQuery dataset. It then invokes BigQuery Data Transfer Service to ingest the data in BigQuery while ensuring the compliance and data annotation are successfully performed for the newly ingested BigQuery data.

\subsubsection{Cloud Dataflow}
Our framework also supports cloud-native solutions to ingest data. We make use of Cloud Dataflow for an easy way to read data from GCS and write to a specific BigQuery dataset. Using Cloud Dataflow also gives a greater amount of flexibility to the users and services to make use of GCP's native APIs and binaries. Our framework requires all data ingestion services to be run from Compute projects since our strict access control policies only allow storing data in the storage projects. As a part of Twitter's Unified Data Processing (UDP) project to migrate analytics jobs to the cloud, Cloud Dataflow plays a vital role in enabling large-scale data analysis on the data stored in the storage projects \cite{bib12} \cite{bib13}.

\begin{table}[h]
\caption{Comparing Data Ingestion Tools}\label{tab:compare-tools}%
\begin{tabular}{@{}llll@{}}
\toprule
Feature & BigQuery Blaster  & BID & Cloud Dataflow\\
\midrule
Supported data formats    & \texttt{Parquet}, \texttt{LZO}   & \texttt{Parquet}, \texttt{CSV}, \texttt{TSV}  & \texttt{LZO}  \\
Data idempotency    & Yes   & Yes  & No  \\
Requires BigQuery Slots    & Yes   & Yes  & Yes  \\
Supports Data Backfill    & Yes   & Yes  & Yes  \\
Partitioned dataset support    & Yes   & Yes  & No  \\
\botrule
\end{tabular}
\end{table}

\subsection{Big Data Analytics}
Since we separated the storage and compute projects, our framework makes it seamless for the services to perform big data analytics on top of BigQuery data. The teams at Twitter heavily use machine learning to analyze the data and perform prediction analysis, BigQuery is a perfect fit for those use cases. Native support for machine learning in BigQuery enables at-scale big data analytics using BigQuery ML \cite{bib14}. BigQuery ML (BQML) enables users to create and execute machine learning models in BigQuery using SQL queries. The services and users can train, evaluate, and even serve models natively from their compute projects while reading the data stored in BigQuery storage projects. Models can be created in BigQuery using the \texttt{CREATE MODEL} statement. This single statement handles the gathering of the training data that is done via a \texttt{SELECT} statement, pre-processing the data, creating the model, training the model, and optionally evaluating the model against validation data. Once the model has been trained, its evaluation and training statistics can be viewed directly in the BigQuery UI, such as the receiver operating characteristic (ROC) curve and the confusion matrix. \\

\begin{verbatim}
CREATE MODEL `mydataset.mymodel'
OPTIONS (MODEL_TYPE=`LINEAR_REG')
AS
SELECT * FROM twitter-my-service-bq-project
\end{verbatim}

\hfill

\noindent Additionally, our framework also seamlessly supports big data analytics using TensorFlow Extended (TFX). TFX is an end-to-end open-source platform for deploying production ML pipelines. A TFX pipeline is a sequence of components that implement an ML pipeline that is specifically designed for scalable, high-performance machine learning tasks \cite{bib15}. Using standard and custom TFX components, you can ingest BigQuery data stored inside the storage projects into a TFX pipeline, upload data to BigQuery datasets, deploy a trained model to BigQuery ML, create a model in BigQuery ML and export it for use in TFX components, and run arbitrary queries. Twitter heavily relies on machine learning-based big data analytics, such as TFX \cite{bib16}, to improve and roll out new features.

\subsection{Slot Management}
Since the big data analytics operations performed over the BigQuery data are compute operations, they need CPU time to perform these operations. CPU time varies based on the type of query, the scale of the query, and data consumed by the machine learning modeling that is run on top of the data stored in BigQuery. A Slot is a unit of computational capacity that may include a combination of RAM and CPU \cite{bib17}. Multiple slots can be classified into the ``buckets" called Reservations. Our framework also makes it possible for the users to consume slots based on their usage and exclusivity of the service. We create two types of slot reservations viz., default and dedicated slot reservations. To start with all the projects that run big data analytics jobs or query data from BigQuery storage projects are added to the default reservation. For example, if we have 100000 slots available in a default reservation, all projects must share this compute capacity. This begets problems like "Noisy Neighbors" where a large analytics job may consume a major chunk of the default reservation keeping other jobs starved. To alleviate this problem, our framework also implements a dedicated reservation where an analytics job is guaranteed a dedicated compute capacity. This dedicated reservation is formed from available default reservation. Thus, if the default reservation has 100000 slots available and if an analytics job ``\texttt{tweet\_analyzer}" requires 30000 slots then for the dedicated reservation is created for the compute project that runs this analytics job. Default reservation, therefore, is left with 70000 slots that are shared among all the compute projects. Although the framework also supports purchasing extra slots real-time, Twitter currently relies on capacity forecasting to buy slots in advance.

\section{Big Data Operations}\label{sec4}
Operations (Ops) is a crucial part of our design framework. It is essential that the users have monitoring and logging support for their big data and analytics job. This framework makes it easier for the admin users and security engineers to audit the operations on the data easily in case of an incident. This section describes how operations are monitored and logged.

\subsection{Monitoring and Alerting}
\begin{figure}
\centering
\begin{tabular}{cc}
 \includegraphics[width=90mm]{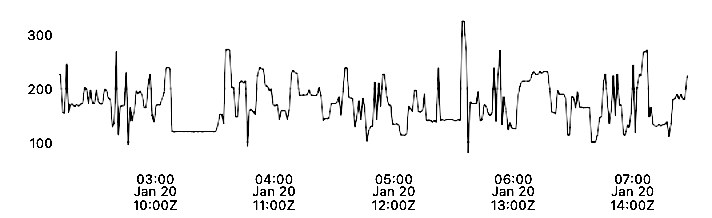} \\
(a) Current Total jobs \\[8pt]
 \includegraphics[width=90mm]{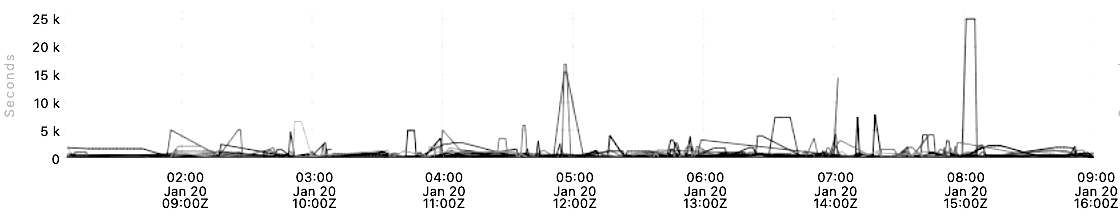} \\
(b) Total Execution Time per Job per Project \\ \\[8pt]
\includegraphics[width=90mm]{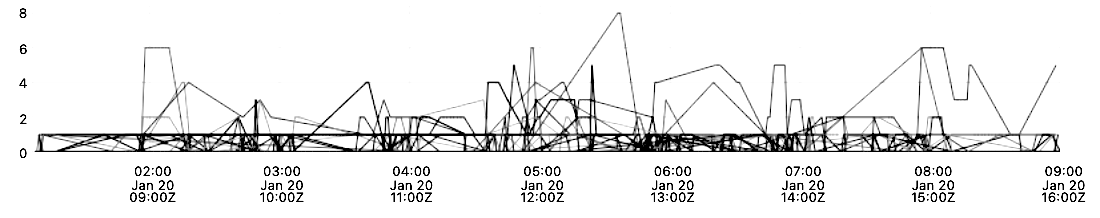} \\[8pt]
{(c) Total Failed Jobs per Project} \\
\end{tabular}
\caption{BigBird Monitoring and Alerting}\label{fig:dashboards}
\end{figure}

To forecast the capacity and usage of the compute and storage quota, we monitor all the jobs, including queries and analytics, that are running at the moment. Since there are hundreds of storage and compute projects, it is not practical to run an "agent" virtual machine in every project tracking the usage and operational statistics. Because of the large scale of the framework, we decided to use a resource at an organization node that will give us bird's eye visibility into all compute and storage projects. We chose to use \texttt{INFORMATION\_SCHEMA} as our basis for fetching all the metrics related to BigQuery usage. \texttt{INFORMATION\_SCHEMA} is a series of views that provide information about BigQuery metrics like the number of datasets, jobs, jobs' timeline, reservation assigned to a job, slots consumed by jobs, and access control. We implemented a central metrics collector agent process that queries \texttt{INFORMATION\_SCHEMA} after a fixed interval, aggregates those metrics based on a query or analytics job, and creates the monitoring dashboards for the data admins to track. Alerts are generated on top of these dashboards and if a metric crosses a certain threshold then the data admins are alerted via a pager software. Fig. \ref{fig:dashboards} showcases a small subset of monitoring dashboards for BigBird. Our framework tracks similar metrics for slots allocation, slot usage, jobs waiting for slot allocation and so on.

\subsection{Audit Logging}
Similar to how compute and storage projects are monitored, we wanted a similar way to generate audit logs for every operation that is performed on the data stored in the storage projects. This includes dataset creation and deletion, data read by a reader group, data written by a shadow service account, and any access control change by an admin account. To achieve this at scale, we set up a logging sink at the organization node with the query that tracks above activities across all the compute and storage projects. The logs are stored inside a BigQuery dataset and persisted as per the compliance requirements.

\subsection{Data Watchdog}
Since data loading in our framework happens in two different phases viz., HDFS to GCS, and GCS to BigQuery, it is essential to have visibility into the delays for data loading. Data Watchdog is a service that analyzes the latencies in jobs that load data from one storage system into another. The monitoring of data pipeline also aids in debugging by providing visibility into what part of the pipeline the data movement is stuck and provide latency visibility across datasets by keeping track of the status of the datasets if the datasets exist in HDFS and GCS, and last time the dataset files were changed. The actual architectural discussion of Data Watchdog is out of scope for this paper. However, it plays an essential part in the monitoring of the overall framework.

\section{Conclusion}\label{sec5}
Our framework provides a scalable way to store more than 300 petabytes of data and enables analytics on the data, including training the machine learning models. Our framework implements different types of projects like compute projects for running compute-based jobs such as querying and analytics, while storage projects are specifically designed for big data storage. Storage projects are further divided into service accounts projects and Scribe log category projects for easy data analysis. The framework also ensures the data is stored securely by following the security principles of Authentication, Authorization, and Accounting (AAA) and Least Privilege Access (LPA) by limiting access to the data to only owner shadow service account while allowing a reader group for identities that want to read the data securely. While the framework is automatically managed end-to-end by an automation service, our future work focuses on empowering the users and services to create their storage projects as per their use case while strictly adhering to the security and privacy, and data protection (PDP) compliance requirements. We hope that our work on big data storage and analytics showcased in this paper would be a guide for industry peers and researchers in understanding how a big data storage and analytics framework can be securely implemented at a large scale. Finally, although our implementation of the framework is in GCP, similar solution can be easily implemented in any major public cloud platform.

\section{Acknowledgements}
The authors thank Abishek Vaithiyanathan, Anju Jha, F. Vivek and Sirushti Murugesan from Twitter, and Ryan McDowell and Vrishali Shah from Google for their contributions to building the scalable Big Data storage and analytics solution at Twitter.

\end{document}